# Gamification Techniques for Raising Cyber Security Awareness

Sam Scholefield
Lynsay A. Shepherd

# Gamification Techniques for Raising Cyber Security Awareness

Sam Scholefield and Lynsay A. Shepherd[0000-0002-1082-1174]

School of Design and Informatics, Abertay University,
Dundee, DD1 1HG, United Kingdom

lynsay.shepherd@abertay.ac.uk

**Abstract.** Due to the prevalence of online services in modern society, such as internet banking and social media, it is important for users to have an understanding of basic security measures in order to keep themselves safe online. However, users often do not know how to make their online interactions secure, which demonstrates an educational need in this area. Gamification has grown in popularity in recent years and has been used to teach people about a range of subjects. This paper presents an exploratory study investigating the use of gamification techniques to educate average users about password security, with the aim of raising overall security awareness. To explore the impact of such techniques, a role-playing quiz application (RPG) was developed for the Android platform to educate users about password security. Results gained from the work highlighted that users enjoyed learning via the use of the password application, and felt they benefitted from the inclusion of gamification techniques. Future work seeks to expand the prototype into a full solution, covering a range of security awareness issues.

**Keywords:** Gamification, games-based learning, security awareness, usable security, human-centered cyber security.

## 1  Introduction

Society has become increasingly reliant on the Internet for banking, and e-commerce. Typical transactions involve the exchange of personal information such as home addresses, and credit card details. Despite the introduction of biometric authentication mechanisms such as fingerprint-based systems [1], passwords continue to be the primary authentication mechanism for accessing such services, therefore it is important to ensure users remain secure online whilst using passwords. The aim of the research presented in this paper is to raise security awareness and improve password hygiene via the use of gamification techniques.

The following sections of the paper will outline the need to improve end-user security awareness, focusing on the topic of password security. Gamification techniques and their application in the context of the learning environment will also be explored before linking these to the domain of security awareness. Subsequently an overview of the Android-based role-playing quiz application (RPG) is provided. Finally, results will be presented and discussed, allowing conclusions to be drawn as to the usefulness of this approach.

## 1.1 Raising End-User Security Awareness

When browsing the web, there are many ways in which users may potentially place themselves at risk. These can include interacting with poorly coded websites, creating weak passwords, and downloading data from websites containing malicious files [2] There are a number of methods which have been used to raise end-user security awareness when engaging in online transactions, from contextual affective feedback presented in a web browser [3-4], to visualizing privacy policies [5], and phishing awareness applications [6]. Owing to the ubiquity of passwords, this work focusses on security awareness tools developed to improve password security.

Users often find the creation and retention of strong, secure passwords to be problematic [7-8], and a number of studies have been conducted to address the issue of security awareness regarding passwords.

A common method of raising password security awareness has involved the use of password meters which are typically placed next to forms on a web page to give users a general indication of password strength. Though meters are widely used, the way in which they measure strength can be poor, meaning trivial passwords can be shown as "safe" [9]. Research has shown additional factors must be considered when using password meters, for example, work by Egelman et al. [10] explored if meters had an impact upon the password created, i.e. if the meter assisted the user in creating a strong password. Results showed that password strength was related to whether the participant felt their account was important, as opposed to the information provided by the password meter. Meters may not necessarily have an impact on raising awareness of creating secure passwords suggesting alternative solutions are needed.

Ciampa [11] also performed a study to explore the impact of different password strength meters, investigating if feedback prompted participants to create stronger passwords. In the study, participants were asked to record four passwords they may use online for accounts. Subsequently, they had to visit websites which offered password strength checking services. Participants also had to record if the password strength checks encouraged them to change their passwords. Results from this experiment showed that *"any feedback mechanism can influence users to create passwords with higher entropy"*. This suggests that user behaviour can be influenced by encouraging users to reflect on their password strength.

Fear appeals are another potential method of raising security awareness. Fear appeals have been described as *"persuasive messages designed to scare people by describing the terrible things that will happen to them if they do not do what the message recommends"* [12]. Vance et al. [13] explored the concept of fear appeals in relation to password security. In this work, participants were asked to register for an account, and the password strength chosen was observed. Multiple groups of participants were used in the study: one group were given no guidance as to how to create a strong password; the static fear appeal treatment group received security information that did not change on user input; another group received an interactive password meter; and a final group received an interactive fear appeal treatment which provided security guidance that updated on user input. Results showed that the interactive fear appeal treatment performed better in terms of choosing stronger passwords, and that such an approach may aid in raising end-user security awareness.

Although previous research has highlighted a number of attempts to raise end-user password security, the prevalence of issues related to password hygiene suggests these are not working. This indicates the need for more effective ways of conveying password security information to the end-user.

### 1.2 Gamification Techniques and Applications

Gamification can be defined as *"the application of gaming mechanics to non-gaming contexts with the aim of inducing engagement and raising levels of motivation"* [14] and aspects of this can be applied to keep a user engaged in learning. Work by Marczewski [15] cataloged the number and types of mechanics which can be used in the process of developing a gamification-based solution. To date, the work has identified 52 mechanisms which can be used, ranging from signposting (preventing users from becoming lost within an application), to providing users with challenges and physical rewards.

Various gamification techniques have also been discussed by Zichermann and Cunningham [16], who explored the concept of a rewards system which can apply to different contexts, known as SAPS (Status, Access, Power, and Stuff). SAPS utilises gamification in the delivery of rewards. Status is derived from how the user performs or compares to their peers. The mechanism of a leaderboard is one method of integrating status into a gamified application as it allows users to compete against each other. Access can be implemented via the use of a loyalty scheme, encouraging users to remain engaged. A notion of power can be achieved by rewarding a user with moderator duties. Finally, the authors explore the category of "stuff" whereby free rewards are given, providing users with an incentive to continue using a particular application or platform.

Several of these gamification techniques have previously been used in educational games such as Duolingo [17] (for learning new languages), and ClassDojo [18] (for parents and teachers to help teach developmental skills to children).

Gamification in education has also been applied to University level courses. Research conducted by Ibanez, Di-Serio and Delgado-Kloos [19] presented the results of a study in which gamified learning activities were used to teach introductory C-programming at undergraduate level. By using a combination of rewards such as points and badges, and allowing students to show their social standing via the use of a leaderboard, the implementation of gamification in this scenario improved knowledge acquisition. However, gamification did not work for all students, whereby some reached one hundred points within the learning activity and stopped playing rather continuing to engage with additional tasks.

Similar work has been carried out at University level by O'Donovan, Gain and Marais [20] drawing similar conclusions, observing that their *"approach to gamification is effective in a university setting"*. Again, they raise similar issues to Ibanez, Di-Serio and Delgado-Kloos [19], noting that gamification must be implemented with careful planning, to ensure it is beneficial.

Given the success of gamification in an educational context, it seems reasonable to suggest that these concepts have the potential to apply to other domains, such as raising security awareness.

### 1.3 Gamification and Security Awareness

A number of cyber security-based games have previously been developed to educate users. Many of these games have been aimed primarily at children and young people, such as the Webonauts Internet Academy, an online game designed to educate children about online etiquette [21]. In this game, users travel around space, visiting different planets, learning to deal with different behaviours exhibited on each of them. These skills are synonymous with behaviour on the Internet.

Another educational game is the Cybersecurity Lab, designed to teach young people basic cyber security skills [22]. In this scenario, the user assumes the role of a Chief Technology Officer at a social media company who must defend the application against a number of attacks. Though the game is designed to provide a level of security knowledge, the educator guide for the game suggests it will take 75 minutes to play through [23], indicating this is a self-contained game which does not promote continuous engagement and ongoing development of cyber security skills. A full browser-based RPG game for children in which they have to save the world from a password crisis [24] was also released.

The role of gamification and cyber security training has also been explored with high school students [25]. Funded by the National Security Agency, and the National Science Foundation, a number of summer camps (named GenCyber) were run in the USA to raise awareness of cyber security, and to encourage interest in computing, covering topics such as secure online behaviour, and social engineering.

Training games in this domain have also been made available for specialist fields, such as law enforcement. Research has been conducted into the use of serious games (i.e. games which are not solely designed for entertainment purposes) examining how these can be used to deliver cybercrime training for law enforcement officers attending a crime scene [26]. A similar game has been made available for Board Members of organisations. Pwc developed Game of Threats [27] which aims to teach the top level of an organisation how to handle a cyber incident.

Pertaining to mobile devices, an Android application targeted towards the general public, called NoPhish [28] was developed to assist users in identifying phishing links. The game consists of multiple levels where users are presented with a URL and are asked to determine its safety. In a subsequent evaluation, participants gave significantly more correct answers when asked about phishing, suggesting this type of application raised their security awareness. A follow-up study was conducted five months later which showed participants still performed well when asking about phishing links however, their overall performance decreased, which suggests issues with retention.

Though many of these games have been created to appeal to children, or specialists, the ubiquity of internet services means that the general population will require security awareness training. The majority of these games are also browser-based, however given the level of worldwide mobile phone ownership [29], it would be beneficial to have a security awareness game application which would appeal to this platform. By creating a level-based approach, this would break learning down into smaller sections, thus ensuring consistency with differing learning attention spans of various age groups [30]. As new cyber security threats are developed, updates could be pushed to the application, keeping end-user knowledge relevant.

Given these factors, the development of a mobile-based application focusing on raising password security awareness has the potential to be an effective tool to help users. In the following section, the methodology behind the research is outlined, explaining how a simplistic password security quiz was turned into a prototype security awareness application via the use of gamification.

## 2  Methodology

As part of an exploratory study, a Unity role-playing quiz application (RPG) was developed for the Android platform to educate users about password security. With a market share of approximately 75% [31], the Android platform was chosen for the development of the prototype application because it would have the ability to reach a larger target audience in comparison to the iOS platform. Similarly, developing the application in Unity provided a number of advantages. Unity is a multi-platform game engine, and the associated Asset Store allows developers to download free and paid-for assets for use in applications created.

On opening the application (Fig. 1), the end-user is presented with 2 characters on the screen: One is a golden knight (the end-user), the other is a dark knight (the character the end-user is fighting) [32]. The application contains questions related to password security, designed to educate the end-user. These questions cover topics such as choosing a strong password, avoiding the use of commonly used passwords, and practicing good password hygiene. If the end-user answers correctly, the dark knight loses health points. If the user is incorrect, the golden knight loses health points. This continues until one character defeats the other, educating the user regarding password security in the process.

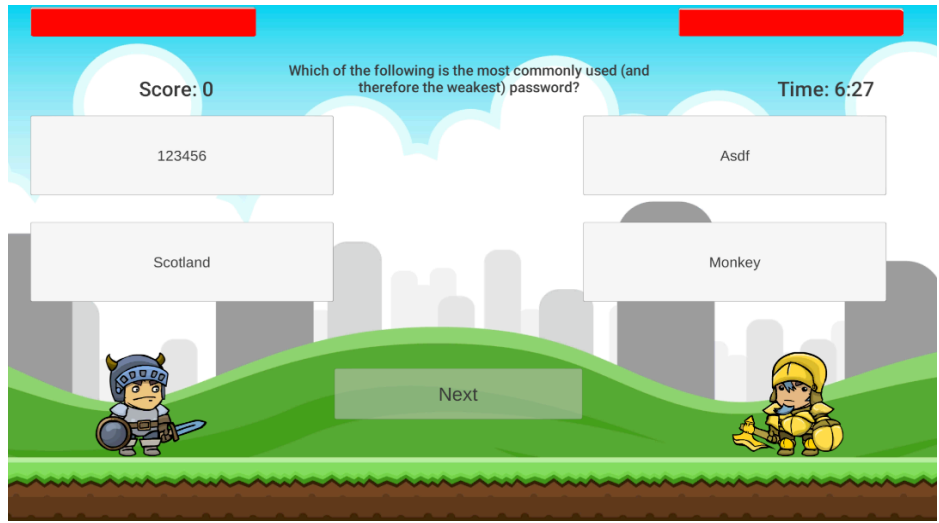

**Fig. 1.** Screenshot from the Android application

The underlying application is a simplistic multiple-choice quiz. However, aspects of gamification were included, with the goal of motivating users to learn and progress. These features were chosen owning to their suitability to integrate into the context of a quiz-based game, enhancing the experience. Gamification features integrated include a specific theme (RPG-style game with characters), on-screen progress/feedback (the health bar per character), time pressure (timer), consequences (if the user is incorrect, they lose health points), and competition (by means of a leaderboard) [15].

To evaluate the potential impact of the application, 17 participants over the age of 18 years were recruited for the pilot study. Participants varied in gender, and level of education. During the evaluation phase, participants were instructed to play through the application.

Following this, they were provided with a series of statements which they were asked to rate against a 5-point Likert scale (1- strongly disagree, 5- strongly agree), providing quantitative data. Statements which the participants were asked to rate included *"Your knowledge in computer security is strong"*, *"The password security game helped increase your knowledge on password security"*, *"The password game was enjoyable"*, and *"Gamification is an effective method of teaching computer security"*. Additionally, a qualitative free-form question was asked, to gather general feedback on the application.

## 3 Results and Discussion

Overall results highlighted that participants exhibited positive opinions towards the use of an RPG-style quiz application which will be discussed in detail, including quantitative data from the Likert-based questions, and qualitative data from the free-form questions.

### 3.1 Quantitative Data

When participants were asked to rate the statement *"Your knowledge in computer security is strong"* using a 5-point Likert scale (1- strongly disagree, 5- strongly agree), only 3 of the participants agreed, or strongly agreed with this assertion, as shown in Fig. 2 (*n=17, mode=2, mean=2.7*).

Therefore, this indicates the majority of participants did not consider themselves to have a good understanding of computer security. This suggests there is a real educational need in this area.

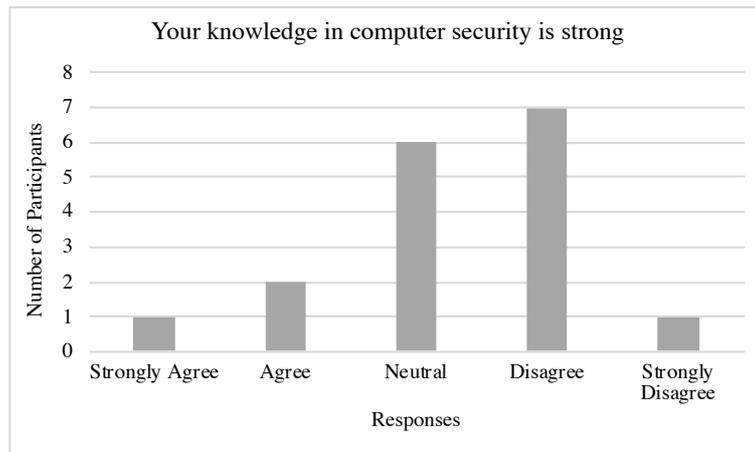

**Fig. 2.** Participants' self-reported knowledge of computer security

When examining the password game and the concept of gamification, results indicated it was well received by participants. Participants were presented with the statement *"The password security game helped increase your knowledge on password security"* (Fig. 3), and the mode indicated the majority agreed (*n=17, mode=4, mean=2.52*).

Although participants *felt* their knowledge of password security increased, this needs further investigation with a longitudinal study. When attempting to raise security awareness, there are typically a number of issues users experience, namely long-term retention, long term behavioural change and security fatigue. The term *"security fatigue"* is linked to security awareness, highlighting that even though there are programmes to educate people about security, people may still fail to engage with the good practice they have been taught [33]. Essentially, users can tire of being bombarded with security information and may reject the advice they have been given [34].

A longitudinal study comparable to work conducted by Canova et al. [28] would allow the long-term impact of the application to be assessed. After using the password security application for a specified period of time, the participants would complete a questionnaire to assess knowledge gained. A similar questionnaire would be given again some months later. Results of the two questionnaires would be compared to establish if playing the password application had a long-term impact on security knowledge.

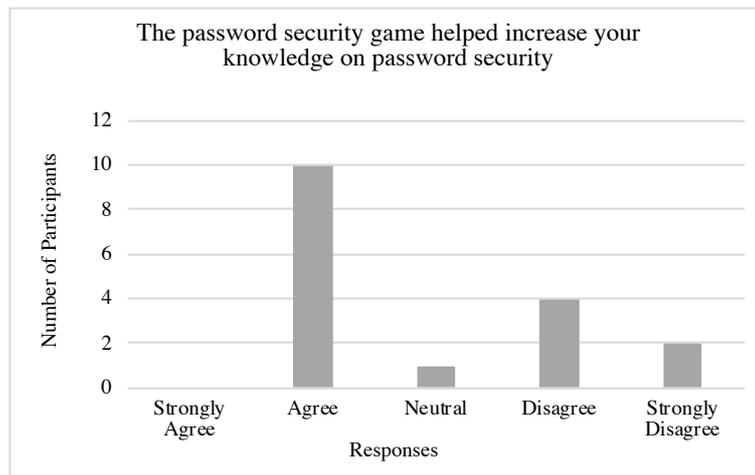

**Fig. 3.** Participants' response when asked if the password security game helped increase their knowledge of password security

Furthermore, the majority of participants agreed with the statement that *"The password game was enjoyable"* (shown in Fig. 4, *n=17, mode=4, mean=3.65*). Regarding the final Likert scale-based statement, *"Gamification is an effective method of teaching computer security"*, again the mode indicated that the majority of participants agreed (shown in Fig. 5, *n=17, mode=4, mean=4.18*).

The values given by participants when asked if they felt gamification was effective are in line with previous research. Canova et al.'s work [28] on developing a phishing awareness game showed that it helped users learn about phishing attacks they may fall victim to. The success of gamification has also been exhibited when teaching a games-development course at undergraduate level [20].

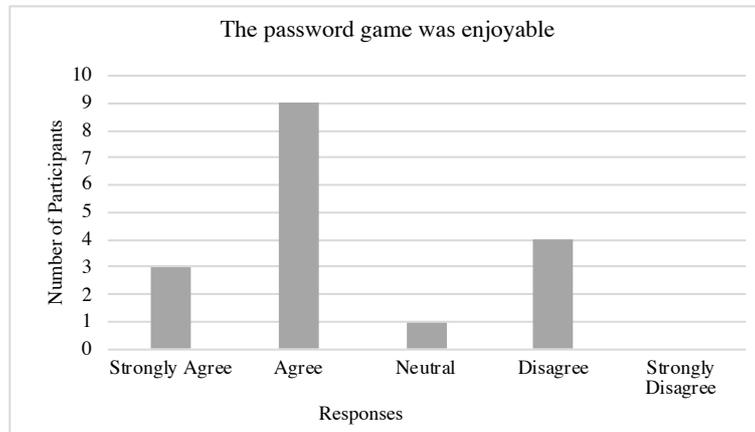

**Fig. 4.** Participants' response when asked if the password game was enjoyable

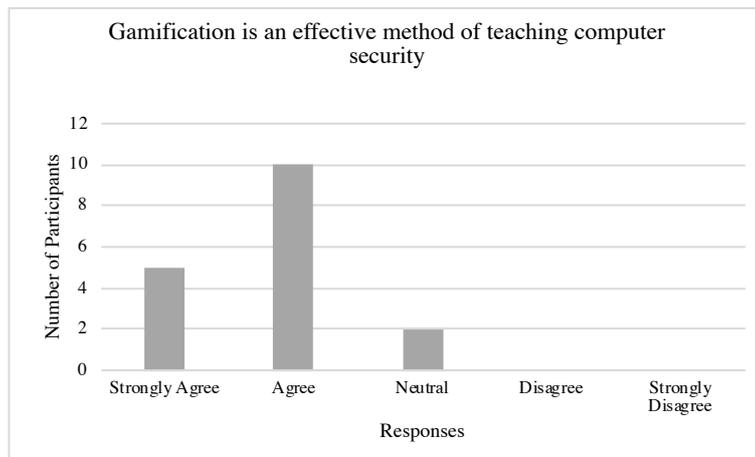

**Fig. 5.** Participants' response when asked if gamification was an effective method of teaching security

However, despite these positive findings, the use of gamification may need to be considered carefully regarding impact. Work by Domínguez et al. [35] conducted a study using gamification, providing exercises on an e-learning platform for a general ICT class, some of which were gamified. Results found that students who engaged with gamified content performed poorly on written tasks, but strongly on practical tasks. This suggests that whilst gamification may be effective, it must be used within the right context.

### 3.2 Qualitative Data

Participants were asked to give an indication as to whether they enjoyed/disliked using the password application. Generally, participants who enjoyed using the application submitted favourable comments relating to the aesthetics of the game, and the implementation of characters. Additionally, these participants claimed it was a fun and interesting method of learning about password security. Those who disliked using the application indicated this was due to a lack of feedback provided, expressing this would have helped them learn from their errors.

**Feedback Provided**

Several of the participants pointed to the lack of feedback provided in the prototype game with statements such as:

– *"Maybe explaining why the questions were wrong would have helped?"*

– *"Fun but wanted answers to wrong questions"*

The need to include feedback whilst utilising gamification is concurrent with work by Ibanez, Di-Serio, and Delgado-Kloos [19], stating that feedback is required, particularly in an educational setting as it prevents people from becoming confused about the current task.

The use of specific words may influence user behaviour when providing feedback, for example, in work by Ur et al. [36] when a password was described as "weak", this prompted users to try and create a stronger password. Text-based feedback is seen to be an appropriate method of delivering feedback to the end-user as it is a more direct way of communicating [37]. When applying feedback in the password application, careful consideration must be given as to how this is implemented, ensuring it is helpful to the user.

**Gamification Elements**

As discussed in the methodology, a number of elements of gamification were included within the application. These comprised of an overarching theme, on-screen feedback, time pressure, consequences, and competition.

*Leaderboards*

Leaderboards are commonly used when introducing the concept of gamification [16][20]. Participants in this exploratory study did not consider the leaderboard which had been implemented to be an important factor in their enjoyment of the game, and instead focussed on other elements.

*Theme*

Many of participants commented on the theme of the application (medieval RPG), with particular comments relating the characters used [32].

— *"The characters were fun to look at, was like playing an RPG"*

— *"I enjoyed the character animations"*

Participants also generally expressed that the overall application was fun, indicating the concept was an interesting way to learn about security.

— *"It looked nice and felt like I was playing a game"*

— *"Learnt information about security in an enjoyable way"*

— *"Was an interesting way to ask questions"*

This result is of interest owing to previous literature establishing how people learn, and the impact which particular themes can have upon the learner. Work by Parker and Lepper [38] examined the use of fantasy contexts in relation to the way children learned to use the Logo programming language. It was found that the use of fantasy contexts such as scenarios involving pirates and detectives motivated children to learn. Given that the password application makes use of a medieval theme, this has the potential to lead to a similar effect. However, others have debated the level of knowledge gained from what is referred to as *"edutainment media"* – games or films which are designed to be fun and educational, and have highlighted the need for longitudinal evaluative studies to be conducted [39].

Overall, qualitative data gained from participants who used the application revealed that, whilst they liked the concept of the application, they felt there were several features missing from the prototype which would have helped their understanding. One such issue was the lack of feedback provided during the game. Ultimately, comments gained from the evaluation were useful for informing how the research work will develop in the future. Before developing a full version of a gamified security awareness application, it is important to consider human-centered design changes following responses from participants.

### 3.3 Limitations

The research contained a number of limitations. This was an exploratory study, with a small sample size, and no long-term evaluation regarding the retention of knowledge in relation to password security was conducted.

Gender differences also need to be explored as this demographic was not included in the participant questionnaire. Previous work in the field has identified that males find game-based learning more enjoyable than females [25]. This may indicate that the password security application may need to be modified to ensure it is consumed by the maximum number of target users.

## 4 Conclusion and Future Work

To conclude, the RPG-style application was viewed positively by participants. The results indicated that participants enjoyed playing this type of application, and they suggested it increased their knowledge on password security. Additionally, participants felt gamification was a useful method of raising security awareness. Owing to the positive results derived from this exploratory study, future work seeks to develop the prototype application into a full security awareness application, covering a range of topics including phishing and information sharing. This will also allow a longitudinal study to be developed to compare knowledge gained from the security awareness application against real-word security practices exhibited by end-users.

Some users highlighted that the application still seemed like a quiz, despite the inclusion of several gamification elements. To overcome this issue, placing the quiz within the context of an overarching storyline may make the application more immersive [15].

Finally, future work seeks to adapt the application to ensure it appeals to varying age ranges, such as children and the elderly, helping them learn about password security in a fun, yet effective manner.